\documentclass[12pt]{article}

\usepackage[english]{babel}
\usepackage{amsfonts}
\usepackage{amssymb}
\usepackage{graphicx}
\newcommand{\be}{\begin{equation}}
\newcommand{\ee}{\end{equation}}
\newcommand{\bea}{\begin{eqnarray}}
\newcommand{\eea}{\end{eqnarray}}

\begin{document}

\begin{titlepage}

\begin{flushright}
 \end{flushright}

\bigskip

\begin{center}

{\Large{\bf{ Semiclassical zero-temperature corrections \\ to
Schwarzschild spacetime and holography}}}

\bigskip
\bigskip\bigskip
 A.  Fabbri$^a$,\footnote{afabbri@ific.uv.es} S.
Farese$^a$,\footnote{farese@ific.uv.es}
 J. Navarro-Salas$^a$,\footnote{jnavarro@ific.uv.es} G. J. Olmo$^{b}$,\footnote{olmoalba@uwm.edu}
 and H. Sanchis-Alepuz$^a$,\footnote{helios.sanchis@ific.uv.es}

\end{center}

\bigskip%

\footnotesize \noindent {\it a) Departamento de F\'{\i}sica
Te\'orica and
    IFIC, Centro Mixto Universidad de Valencia-CSIC.
    Facultad de F\'{\i}sica, Universidad de Valencia,
        Burjassot-46100, Valencia, Spain.}\\
{\it b) Department of Physics, University of Wisconsin-Milwaukee,
P.O. Box 413, Milwaukee, Wisconsin, 53201 USA}

\bigskip

\bigskip

\begin{center}
{\bf Abstract}
\end{center}

Motivated by  the quest for  black holes in $AdS$ braneworlds, and
in particular by the holographic conjecture relating $5D$
classical bulk solutions with $4D$ quantum corrected ones, we
numerically solve the semiclassical Einstein equations
(backreaction equations) with matter fields in the (zero
temperature) Boulware vacuum state. In the absence of an exact
analytical expression for $\langle T_{\mu \nu}\rangle$ in four
dimensions we work within the s-wave approximation. Our results
show that the quantum corrected solution is very similar to
Schwarzschild till very close to the horizon, but then a bouncing
surface for the radial function appears which prevents the
formation of an event horizon. We also analyze the behavior of the
geometry beyond the bounce, where a curvature singularity arises.
In the dual theory, this indicates that the corresponding $5D$
static classical braneworld solution is not a black hole but
rather a naked singularity.

\bigskip


\end{titlepage}

\newpage

\section{Introduction}
The study of quantum effects in black hole spacetimes comes back
to the early seventies, when Hawking  discovered \cite{hawk1} that
black holes evaporate by emission of thermal radiation (see also
\cite{parker, wald}). This result generated enormous interest in
the subject, especially after Hawking himself speculated
\cite{hawk2} that the evaporation process will lead to the
disappearance of the black hole and the information about its
formation will be lost forever. This is a radical conclusion, as
it implies that in the quantum theory the whole process of black
hole formation and evaporation is nonunitary.

It is clear, however, that the approximation considered to derive
this result, i.e. the quantization of matter fields in the fixed
classical background describing the formation of a Schwarzschild
black hole, and even the framework used, the semiclassical theory
of gravity (see, for instance, \cite{birreldavies, frolovnovikov,
icp05}), cannot lead to a reliable resolution of this paradox. At
a certain point during the evolution the quantum effects will
backreact and modify significantly the background geometry, which
therefore cannot be considered as fixed, nor evolved in a
quasi-static approximation. Moreover, once the black hole has
reached the Planck size quantum gravitational effects will become
important and cannot be neglected anymore.\footnote{To find a way
out to this problem one usually argues that quantum gravitational
effects should always be negligible compared to those due to a
large number $N$ of matter fields.} Thus it is no wonder that
still today Hawking's provocation continues to raise much debate
and although most of the people do not want to give up unitarity
\cite{unit} (and between them, remarkably, now Hawking himself
\cite{hawk3}), a definitive answer on whether and how information
is recovered in black hole evaporation is still lacking.

To take into account self-consistently the backreaction effects
one needs to solve exactly the semiclassical Einstein equations
\be \label{semeqs0} G_{\mu\nu}(g_{\alpha\beta})= \frac{8\pi
G}{c^4} \langle\Psi|T_{\mu\nu}(g_{\alpha\beta})|\Psi\rangle \ee
for the metric $g_{\alpha\beta}$, where the quantity on the right
hand side represents the expectation value of the stress-energy
tensor operator of the matter fields in a suitable quantum state
$|\Psi\rangle$.

In the fixed Schwarzschild background \be \label{schw}
ds^2=g_{\mu\nu}^{schw}dx^\mu dx^\nu=-(1-r_S/r)c^2dt^2+
\frac{dr^2}{(1-r_S/r)}+r^2d\Omega^2 \ , \ee where $r_S= 2GM/c^2$
is the Schwarzschild radius, three inequivalent quantum vacuum
states can be defined. The first is the Boulware state $|B\rangle$
\cite{boulware}, probably the most natural one, constructed by
requiring that in the asymptotic region, where the metric becomes
minkowskian, it reduces to Minkowski ground state $|M\rangle$. It
has the property that $\langle B|T_{\mu\nu}|B\rangle$ vanishes
asymptotically, but the drawback is a strong divergence at the
horizon $r=r_S$ \cite{chrisful}. One can circumvent this
difficulty by introducing a new quantum vacuum state $|H\rangle$,
called the Hartle-Hawking state \cite{harhaw}, such that $\langle
H|T_{\mu\nu}|H\rangle$ is regular at the horizon. However one pays
a price, i.e. this stress tensor is nonvanishing at large $r$ and
describes thermal radiation at the Hawking temperature \be
T=\frac{\hbar c^3}{G k_B}\frac{1}{8\pi M}\ , \ee where $k_B$ is
Boltzman's constant. The associated physical situation is that of
a black hole in a cavity, in thermal equilibrium with its own
radiation. The third possibility, the Unruh state $|U\rangle$
\cite{unruh}, is constructed in such a way as to reproduce the
late time behaviour of the quantum matter fields in the classical
background of a collapsing star forming a black hole. By requiring
that no particles are present at infinity in the past and that
$\langle U|T_{\mu\nu}|U\rangle$ is regular at the future horizon
one finds, as a consequence,  an outgoing flux of thermal
radiation (the Hawking flux) in the asymptotic future.

By inserting the fixed background expression 
$\langle\Psi|T_{\mu\nu}(g_{\alpha\beta}^{schw})|\Psi\rangle$ in
the right hand side of Eqs. (\ref{semeqs0}) one can solve
perturbatively the backreaction equations at $O(\hbar)$ to find
the first order quantum corrections $\delta g_{\alpha\beta}$ to
the Schwarzschild metric, i.e.
$g_{\alpha\beta}=g_{\alpha\beta}^{schw}+ \delta g_{\alpha\beta}$.
This is a good approximation to the full solution of
(\ref{semeqs0}) only when $\delta g_{\alpha\beta}\ll
g_{\alpha\beta}^{schw}$, i.e. when the quantum terms are small
compared to the background.\footnote{The natural coordinates
to be used in this context are the Schwarzschild ones given in 
(\ref{schw}). In more general terms, this approximation scheme
is valid whenever the quantum terms generate a correction to
the curvature which is small compared to its background value.} 
In the case of large mass being the
Hawking temperature very small this condition is satisfied, for an
evaporating black hole, for most of its lifetime, but eventually
at the late stages of the evaporation one faces the problems
mentioned above. Turning to the static configurations, this is a
good approximation in the Hartle-Hawking case since $\langle
H|T_{\mu\nu}|H\rangle$ never gets large, while for Boulware it is
certainly valid at infinity but not at the horizon due to the
divergence of the quantum stress tensor there. The problem of
understanding how, in this case, the quantum effects modify the
structure of the classical horizon is not an easy one. One usually
disregards this question since  Boulware state describes the
vacuum polarization around a static star whose radius is bigger
than $r_S$, and therefore this divergence is not physically
relevant.

The motivation for our work comes from braneworld physics, where
much work is being done on the search for black hole solutions in
the Randall-Sundrum model RS2 \cite{ransun}. This is technically a
very involved situation and nobody so far has achieved the goal of
finding a five-dimensional solution describing a black hole
localized on the brane \cite{molti}. An interesting physical
interpretation of this fact comes from the application of the
holographic AdS/CFT correspondence \cite{malda}, for which
classical 5D bulk solutions are dual to 4D self-consistent
semiclassical configurations where gravity is coupled to quantum
matter fields \cite{taefk}. If this is so then a classical static
5D `` black hole''  will be mapped to a 4D static solution of the
backreaction equations. Staticity naturally selects the
Hartle-Hawking and Boulware states. If, in addition, we require
that the 4D configuration be asymptotically flat then the choice
must be the Boulware state. That this is the correct choice is
supported by the fact that for large radius the corrections to the
Newtonian potential in this state match those calculated
classically in 5D \cite{abf}.


In order to solve the backreaction equations in the Boulware
vacuum the exact expression of $\langle
B|T_{\mu\nu}(g_{\alpha\beta})|B\rangle$ for an arbitrary geometry
is needed. No such expression exists in four dimensions (an
analytic approximation for static spherically symmetric spacetimes
has however been developed in \cite{ahs}). The situation greatly
improves if one restricts to the s-wave approximation for
spherically symmetric backgrounds. Within this context, in section
2 we review the classical and semiclassical theory of gravity
coupled to a massless and minimally coupled scalar field. Since in
this case the expression of $\langle
B|T_{\mu\nu}(g_{\alpha\beta})|B\rangle$ is available, we end the
section by writing down the relevant backreaction equations. These
equations cannot be exactly solved analytically, and we approach
the problem in two steps. First, in section 3, we consider the
equations arising in the Polyakov theory, which can be derived
from ours using a near-horizon approximation for the scalar field
(this amounts to neglecting backscattering effects in the
propagation of the matter fields). These equations are decoupled,
in the sense that one can derive an equation relating the
conformal factor of the metric $\rho$ ($ds^2 =
e^{2\rho}(-c^2dt^2+dx^2)+r^2d\Omega^2$) as a function of the
radius $r$ only (or $\phi$ or $z$ through the definitions
$r=r_0e^{-\phi}=r_0 z$ according to the convenience) which is then
integrated numerically and from which one can derive the
dependence on the spatial coordinate $\rho=\rho(x)$ and $r=r(x)$.
The quantum corrected solution is very similar to the
Schwarzschild solution from  infinity till very close to the
classical horizon $r=r_S$, where, as expected, big differences
emerge. In particular, there exists a timelike surface $r=r_B$
where the radial function $r$ bounces (i.e. the two-spheres reach
a minimum radius $r_B$ and then they start to increase) and,
beyond it, a null curvature singularity with infinite radius at a
finite affine distance. Armed with these techniques and results we
face, in section 4, the full backreaction equations in the s-wave
approximation, which are much more complicated. The differences
with respect to the previous case are that the bounce  is located
closer to the classical horizon $r=r_S$ and that the curvature
singularity is now timelike and has finite radius. Finally, in
section 5 we summarize our conclusions.

\section{Gravity coupled to a  massless scalar field in the $s$-wave approximation}

It is convenient, in the context of the $s$-wave approximation, to
work with spherically reduced theories. Under the spherically
symmetric ansatz \be \label{sphsymetric} ds_{(4)}^2=ds_{(2)}^2+r^2
d\Omega^2, \ee the Hilbert-Einstein action \be
S^{(4)}_g=\frac{c^3}{16\pi G}\int d^4x\sqrt{-{g^{(4)}}}R^{(4)}\ee
 reduces to \be\label{eq:2DG}
S_g=\frac{c^3}{4G}\int d^2x\sqrt{-g}\left [r^2R+2\left(1+|\nabla
r|^2\right)\right ],\ee where the geometrical quantities refer to
the radial part $ds^2_{(2)}$ of the four-dimensional metric. Note
that the radial variable $r$ here plays the role of a scalar field
with a non-trivial coupling to the radial sector of the metric.
Einstein's equations for spherically symmetric configurations in
vacuum can be rewritten as \bea \frac{2}{\sqrt{-g}}\frac{\delta
S_g}{\delta g^{ab}} &\equiv& \frac{c^3}{4G}\left
[-2r\nabla_{a}\nabla_{b}r +g_{ab}\left( 2r\Box r -1+|\nabla
r|^2\right) \right]=0 \ , \\
\frac{2}{\sqrt{-g}}\frac{\delta S_g}{\delta r } &\equiv&
\frac{c^3}{G}\left [r R-2\Box r \right ]= 0 \ .
\label{eq:eom-Pol-2}\end{eqnarray} The solution of these equations
is the Schwarzschild geometry \bea \label{classicalsolutions}
ds^2_{(2)}&=& -(1-\frac{2GM}{c^2
r})(dt^2-dr^{*2}) \nonumber \\
r^* &=& r + \frac{2GM}{c^2}\ln(1-\frac{2GM}{c^2r}) \ , \eea where
$r^*$ is the so-called ``tortoise'' coordinate.

Turning to the matter sector, let us consider the action for a
minimally coupled massless scalar field (in Gaussian units) \be
\label{a4d}S^{(4)}_{m} = -\frac{1}{8\pi}\int
d^4x\sqrt{-g^{(4)}}(\nabla f)^2
 \ . \ee
In the background $
 ds^2_{(4)}
= g_{ab}dx^adx^b+r^2d\Omega^2 $ the field $f$ can be expanded in
spherical harmonics, of which we pick up only the $s$-wave
component  \be f=f(x^a)\equiv \frac{f_{l=0}}{r}Y_{00} . \ee Under
this assumption, integration of the angular variables in
(\ref{a4d}) leads to
  \be \label{a2dm}S_{m}=-\frac{1}{2}\int
d^2x\sqrt{-g}r^2(\nabla f)^2 \ . \ee Varying this action with
respect to the radial part of the metric we obtain a
two-dimensional stress-energy tensor \be
-\frac{2c}{\sqrt{-g}}\frac{\delta S_m}{\delta g^{ab}} \equiv
T_{ab} \ , \ee which is related to the radial components of the
corresponding four-dimensional one by the relation \be
T^{(4)}_{ab}=\frac{T_{ab}}{4\pi r^2} . \ee Moreover, by varying
(\ref{a2dm}) with respect to $r$ we get the expression for the
angular components of the four-dimensional stress-energy tensor
\be
T^{(4)}_{\theta\theta}=\frac{T^{(4)}_{\varphi\varphi}}{\sin^2\theta}=-\frac{rc}{8\pi\sqrt{-g^{(2)}}}\frac{\delta
S_m}{\delta r} \ . \ee

\subsection{Semiclassical theory}

The advantage of the approximation considered is that in this
case, unlike the full four-dimensional treatment, one can provide
an analytic expression for the expectation values of all
components of the stress-energy tensor. We shall briefly review
the main steps involved. The details can be found in \cite{icp05}.
To this end it is very convenient to parameterize the radial part of the four
dimensional metric in
conformal gauge as \be ds^2_{(2)} = -e^{2\rho}dx^+dx^- , \ee and  moreover
it is also useful to parameterize the radial coordinate as follows
\be r=r_0e^{-\phi} . \ee

One can univocally provide an expression for $\langle
T^{(4)}_{\pm\pm}\rangle$ and $\langle
T^{(4)}_{\theta\theta}\rangle$ by imposing two simple conditions:
\begin{itemize}
\item the covariant conservation laws \be \nabla^{\mu} \langle
T_{\mu\nu}^{(4)}\rangle =0 , \ee which can be rewritten as \be
\nabla^a \langle T_{ab}\rangle=\nabla_b \phi
\frac{1}{\sqrt{-g}}\langle\frac{\delta S_m}{\delta \phi}\rangle
 \ ; \ee

\item at an arbitrary  point $X$ of the spacetime manifold the
expectation values of the quantum stress-energy tensor $\langle
T_{\pm\pm}(x^{\pm}(X))\rangle$ reduce to the normal ordering ones
$\langle :T_{\pm\pm}(x^{\pm}(X)):\rangle$ when using
 a locally inertial frame $\xi^{\alpha}_X$ based on that point  \be
\langle T_{\pm\pm}(\xi^{\alpha}_X(X))\rangle = \langle
:T_{\pm\pm}(\xi^{\alpha}_X(X)):\rangle . \ee
\end{itemize}

These two conditions are strong enough to provide a generic
expression for the expectation values of the stress-energy tensor.
In particular, the breaking of the classical  Weyl symmetry
(meaning that classically $g^{ab}T_{ab}=0$), produces a
non-vanishing trace anomaly which can be derived from the above
conditions. One easily obtains that \be \langle T \rangle =
\frac{\hbar}{24 \pi}( R -6(\nabla \phi)^2+6 \Box \phi) \ . \ee The
full expression for the expectation values of the stress-energy
tensor components, in an arbitrary conformal coordinate system, is

 \bea \label{uno}\langle\Psi |T_{\pm\pm}(x^\pm)|\Psi \rangle
&=&-\frac{\hbar}{12\pi}(\partial_\pm\rho\partial_\pm\rho-\partial^2_\pm\rho)+\frac{\hbar}{2\pi}
(\partial_\pm\rho\partial_\pm\phi+\rho(\partial_\pm\phi)^2)
\nonumber  \\
&+&\langle\Psi
|:T_{\pm\pm}(x^\pm):|\Psi \rangle \ , \\
\label{due} \langle\Psi |T_{+-}(x^\pm)|\Psi \rangle &=&
-\frac{\hbar}{12\pi}(\partial_+\partial_-\rho
+3\partial_+\phi\partial_-\phi-3\partial_+\partial_-\phi) \ , \\
\label{tre} \langle\Psi|\frac{\delta
S_m}{\delta\phi}|\Psi\rangle&=&\langle\Psi|\frac{\delta
S_m}{\delta\phi}|\Psi\rangle_{\rho=0}-\frac{\hbar}{2\pi}(\partial_+\partial_-\rho+\partial_+\rho\partial_-\phi+\partial_
-\rho\partial_+\phi \nonumber
\\&+&2\rho\partial_+\partial_-\phi) \ . \eea

The dependence  on the quantum state is all contained in the three
functions $\langle\Psi|:T_{\pm\pm}:|\Psi\rangle$ and $ \langle
\Psi|\frac{\delta S_m}{\delta \phi}|\Psi\rangle _{\rho=0}$. These
functions are not independent and verify the following relations
\be \label{conslawflatspace}
\partial_{\mp}\langle\Psi|:T_{\pm\pm}:|\Psi\rangle + \partial_{\pm}  \phi \langle \Psi|\frac{\delta
S_m}{\delta \phi}|\Psi \rangle _{\rho=0}- \frac{\hbar}{4\pi
}\partial_{\pm}(\partial_+\phi \partial_-\phi -
\partial_+\partial_-\phi) =0\ . \ee

\subsection{Backreaction equations in the Boulware state}

 In  a dynamical scenario such as black hole evaporation
 it is highly nontrivial to unravel the precise form of the
 state-dependent functions $\langle\Psi|:T_{\pm\pm}:|\Psi\rangle$ and $ \langle
\Psi|\frac{\delta S_m}{\delta \phi}|\Psi\rangle _{\rho=0}$.
However, in this paper we are interested in static configurations,
for which $\rho$ and $\phi$ are functions of the spatial
coordinate $x=(x^+ - x^-)/2$ only, i.e. $\rho=\rho(x)$ and $\phi =
\phi(x)$. This coordinate $x$ reduces, in the classical limit, to
the tortoise coordinate $r^*$ given in (\ref{classicalsolutions}).
For the Boulware state it is natural to impose that \be
\label{condboul} \langle B|:T_{\pm\pm}(t,x):| B\rangle =0\ . \ee
This allows to determine, from Eqs. (\ref{conslawflatspace}), the
function $\langle \Psi|\frac{\delta S_m}{\delta \phi}|\Psi \rangle
_{\rho=0}$  \be \label{condphi}\langle B|\frac{\delta S_m}{\delta
\phi}| B \rangle _{\rho=0}=-\frac{\hbar}{16 \pi}
\frac{(\phi_{x}^2-\phi_{xx})_{x}}{\phi_x} \ , \ee where the index
 $x$ means  derivative with respect to the coordinate $x$. Thus we
have all the ingredients we need to write down the backreaction
equations in the Boulware state, which describe how the
Schwarzschild solution is modified due to pure vacuum polarization
effects
\bea\label{semeqs} \frac{2c}{\sqrt{-g}}\frac{\delta S_g}{\delta
g^{\pm\pm}}&=&\langle\Psi|T_{\pm\pm}|\Psi\rangle \ , \nonumber \\
\frac{2c}{\sqrt{-g}}\frac{\delta S_g}{\delta
g^{+-}}&=&\langle\Psi |T_{+-}(x^\pm)|\Psi \rangle \ , \nonumber \\
-\frac{\delta S_g}{\delta \phi} &=&\langle \Psi|\frac{\delta
S_m}{\delta \phi}|\Psi \rangle  \ , \eea where \bea
\label{2deinsteinsemiclassical1} \frac{2}{\sqrt{-g}}\frac{\delta
S_g}{\delta g^{\pm\pm}} &=& \frac{c^3 r_0^2e^{-2\phi}}{G}\left
(\partial_{\pm}^2 \phi -2
\partial_{\pm} \rho
\partial_{\pm} \phi -(\partial_{\pm}\phi)^2 \right )\ ,   \nonumber \\
 \frac{2}{\sqrt{-g}}\frac{\delta
S_g}{\delta g^{+-}} &=& \frac{c^3 r_0^2e^{-2\phi}}{G} (-\partial_+
\partial_-\phi +2\partial_+\phi \partial_-\phi
 +\frac{1}{4r_0^2}e^{2(\rho+\phi)})
  \ , \nonumber \\
 -\frac{\delta S_g}{\delta \phi} &=& 2\frac{c^3 r_0^2e^{-2\phi}}{G}(\partial_+\partial_-\rho + \partial_+\phi
 \partial_-\phi -\partial_+ \partial_-\phi)
 \ . \eea
It is convenient to  fix the constant scale $r_0$ as follows:
$r_0\equiv \sqrt{\lambda} =
\sqrt{\frac{l_{Planck}^2}{12\pi}}=\sqrt{\frac{\hbar G}{12\pi
c^3}}$.

The static differential equations corresponding to Eqs. (\ref{semeqs})
can then be written as

\begin{eqnarray}
\phi_{xx}-\phi_x^2-2\rho_x\phi_x&=&e^{2\phi}\left[\rho_{xx}-\rho_x^2+6\rho_x\phi_x+6\rho\phi_x^2\right] \label{eq:cg-1}\\
\phi_{xx}-2\phi_x^2+\frac{e^{2(\phi+\rho )}}{\lambda}&=& e^{2\phi}\left[\rho_{xx}-3(\phi_{xx}-\phi_x^2)\right]  \label{eq:cg-2}\\
\phi_{xx}-\phi_x^2-\rho_{xx}&=&e^{2\phi}\left[3\rho_{xx}+6\rho_x\phi_x+6\rho\phi_{xx}+\frac{3}{2}\frac{(\phi_{xx}-\phi_x^2)_x}{\phi_x}
\right]. \ \ \ \label{eq:cg-3}
\end{eqnarray}
To solve these equations we have to add boundary conditions which,
in the present context, are naturally given by imposing that for
very large $r$ the solution approaches the classical one
(\ref{classicalsolutions}), i.e.,  \bea
\label{classicalsolutions2} \rho&=&
\frac{1}{2}\ln(1-\frac{2GM}{c^2
r}) \ ,  \\
\label{classicalsolutions22}r^*\equiv x &=& r +
\frac{2GM}{c^2}\ln(1-\frac{2GM}{c^2r}) \ . \eea We shall
investigate how the relations $\rho=\rho(r)$ and $ r=r(x)$ are
modified by (static) quantum effects.

It is convenient to analyze first what happens in a simplified
context, defined by neglecting the coupling of the scalar field
with the radial function $r$ in the classical matter action
(\ref{a2dm}).

\section{Polyakov theory's approximation}

 In this section we shall study a simplified version of the problem outlined in the previous section.
As already mentioned we shall replace the action (\ref{a2dm}) by a
new one obtained by fixing the radial function $r= r_S= constant$.
We then obtain
  \be \label{classicalmp}
S_{matter}=-\frac{1}{2}\int d^2x \sqrt{-g} |\nabla (r_Sf) |^2 \ .
\ee This approximation is usually motivated by arguing
that, in the vicinity of the classical horizon $r \sim
r_S=2GM/c^2$, the
 wave equation for the scalar field \be \label{waveequations=0}
(-\frac{\partial^2}{\partial t^2}+ \frac{\partial^2}{\partial
{r^*}^2}- V(r))(rf)=0 \ , \ee where $V(r)$ is the $s$-wave
potential \be
\label{spotential}V(r)=(1-\frac{r_S}{r})\frac{r_S}{r^3} \ , \ee
reduces to the two-dimensional free wave equation \be
(-\frac{\partial^2}{\partial t^2}+ \frac{\partial^2}{\partial
{r^*}^2})(r_S f)=0 \ . \ee This latter equation can indeed be
derived by varying the action (\ref{classicalmp}).

The expression for $\langle \Psi|T_{ab}|\Psi\rangle$ can be
derived in a number of different ways. Following the arguments of
subsection 2.1 one arrives at
\bea \label{pol1} \langle\Psi|T_{\pm\pm}|\Psi\rangle &=&
-\frac{\hbar}{12\pi} \left( (\partial_{\pm}\rho)^2
-\partial_{\pm}^2\rho \right) +
\langle\Psi|:T_{\pm\pm}:|\Psi\rangle\ ,  \\ \label{pol2}
\langle\Psi|T_{+-}|\Psi\rangle &=& -\frac{\hbar}{12\pi}
\partial_+\partial_-\rho \ , \eea
which are obtained from  (\ref{uno}), (\ref{due}) by neglecting
the terms depending on $\phi$ (note that in this approximation
$\langle\Psi| \frac{\delta S_m}{\delta \phi}|\Psi\rangle =0$).
Note that the above expressions can  be also obtained from the
effective Polyakov action  \be S_P= - \frac{\hbar}{96 \pi }\int
d^2x \sqrt{-g}R \Box^{-1} R \ . \ee

The dependence on the quantum state is contained in the functions
$\langle\Psi|:T_{\pm\pm}(x^{\pm}):|\Psi\rangle$, which are taken
to be zero in the Boulware vacuum, i.e. \be \label{boulpol}
\langle B|:T_{\pm\pm}(x^{\pm}):| B\rangle =0\ . \ee In the
Schwarzschild background ((\ref{classicalsolutions2}) and
(\ref{classicalsolutions22})) $x^{\pm}$ denote the
Eddington-Finkelstein coordinates and the components of the
quantum stress tensor read \bea \label{schboulpol} \langle
B|T_{\pm\pm}| B\rangle &=& \frac{\hbar}{24\pi} [
-\frac{r_S}{2r^3}+
\frac{3}{8}\frac{r_S^2}{r^4} ]\ , \nonumber \\
\langle B|T_{+-}| B\rangle &=& -\frac{\hbar}{24\pi}
[1-\frac{r_S}{r}] \frac{r_S}{2r^3}\ . \eea At infinity $\langle
B|T_{ab}| B\rangle \to 0$ (where $|B\rangle$ reduces to the
Minkowski ground state $|M\rangle$), while on the horizon $\langle
B|T_{\pm\pm}| B\rangle \to - \hbar c^4/768\pi M^2 G^2$. These
quantities are strongly divergent when expressed in Kruskal
coordinates $U\sim e^{-x^-/2r_S},\ V\sim e^{x^+/2r_S}$ regular on
the future and past horizons ($H^+$ and $H^-$, respectively)\be
\label{div} \langle B|T_{UU}| B\rangle \sim_{H^+} \frac{\langle
B|T_{--}| B\rangle}{(r-r_S)^2}, \ \ \ \langle B|T_{VV}| B\rangle
\sim_{H^-} \frac{\langle B|T_{++}| B\rangle}{(r-r_S)^2}\ .\ee
This, in turn, means that quantum backreaction effects are strong
at the classical horizon $r=r_S$.

The semiclassical equations in the Boulware vacuum can  then be
written as follows
 \bea \label{2deinsteinsemiclassical1}
\frac{c^3 r_0^2e^{-2\phi}}{G}\left (\partial_{\pm}^2 \phi -2
\partial_{\pm} \rho
\partial_{\pm} \phi -(\partial_{\pm}\phi)^2 \right ) &=&  -\frac{\hbar}{12\pi} \left( (\partial_{\pm}\rho)^2
-\partial_{\pm}^2\rho \right) \ \ \ \ \  \\
 \frac{c^3 r_0^2e^{-2\phi}}{G}
(-\partial_+ \partial_-\phi +2\partial_+\phi \partial_-\phi
 +\frac{1}{4r_0^2}e^{2(\rho+\phi)})
 &=& -\frac{\hbar}{12\pi}
\partial_+\partial_-\rho\\
  2\frac{c^3 r_0^2e^{-2\phi}}{G}(\partial_+\partial_-\rho + \partial_+\phi
 \partial_-\phi -\partial_+ \partial_-\phi)&=&0
 \  \eea
and repeating the steps that led to Eqs. (\ref{eq:cg-1}),
 (\ref{eq:cg-2}) and (\ref{eq:cg-3}), we rewrite them as
\begin{eqnarray}
\phi_{xx}-\phi_x^2-2\rho_x\phi_x&=&e^{2\phi}\left(\rho_{xx}-\rho_x^2\right)
\label{eq:cg-i}\ , \\
\phi_{xx}-2\phi_x^2+\frac{e^{2(\phi+\rho )}}{\lambda}&=& e^{2\phi}\rho_{xx}
\label{eq:cg-ii}\ , \\
\phi_{xx}-\phi_x^2-\rho_{xx}&=&0 \label{eq:cg-iii}
 \ . \end{eqnarray}
We note that when the right hand side of the above equations
vanishes,  while keeping finite the quotient $e^{2\phi}/\lambda
\equiv r^{-2}$, we recover the classical equations and therefore
the Schwarzschild solution. Due to the divergent behavior in
(\ref{div}), when $r$ approaches the classical horizon $r_S$
we expect non-trivial
corrections to the classical metric.

\subsection{Decoupling the semiclassical equations}

We shall exploit the fact that Eqs.
(\ref{eq:cg-i}),(\ref{eq:cg-ii}) and (\ref{eq:cg-iii}) do not have
terms that depend explicitly on the variable $x$ and, also, that
(\ref{eq:cg-i}) and (\ref{eq:cg-iii}) are homogeneous differential
equations of order two. This allows to write a decoupled equation
for the function $\rho(\phi)$.  We use the relations
\begin{eqnarray}
\rho_x &=&\dot{\rho }\phi_x \ , \\ \label{roxx}
\rho_{xx}&=&\ddot{\rho}(\phi_x)^2+\dot{\rho}\phi_{xx}\ ,
\end{eqnarray}
where the dot indicates derivative with respect to $\phi$.
Equation (\ref{eq:cg-i}) can be rewritten as
\begin{equation}\label{eq:cg-4}
\phi_{xx}=\frac{1+2\dot{\rho}+ e^{2\phi}[\ddot{\rho }-\dot{\rho
}^2]}{1-e^{2\phi}\dot{\rho }}\phi_x^2 \ . \end{equation}
Moreover, subtracting Eq.(\ref{eq:cg-iii}) from (\ref{eq:cg-i}) we get
\begin{equation}\label{eq:cg-5}
\ddot{\rho }+\dot{\rho }\frac{\phi_{xx}}{\phi_x^2}=
\frac{\dot{\rho} (2- e^{2\phi}\dot{\rho})}{1-e^{2\phi}}\ .
\end{equation}
Equations (\ref{eq:cg-4}) and (\ref{eq:cg-5}) allow to obtain the
desired equation
\begin{equation}\label{eq:cg-6}
\ddot{\rho }=\frac{\dot{\rho }\left[1-2\dot{\rho }+
e^{2\phi}(1-\dot{\rho }+\dot{\rho }^2)\right]}{1-e^{2\phi}}
\end{equation}
or, going back to the radial coordinate $r=r_0e^{-\phi}$,
\begin{equation}\label{eq:cg-7}
\rho_{rr}=-\frac{(1+r\rho_r)\rho_r(2r+ \lambda \rho_r)}{r^2-
\lambda}
 \ . \end{equation}
The equation for $\phi(x)$ can be derived by combining
Eqs. (\ref{eq:cg-ii}) and (\ref{eq:cg-4})  \be
\phi_{x}^2= \frac{e^{2(\phi + \rho)}}{\lambda(1-2\dot{\rho} +
e^{2\phi}\dot{\rho}^2)} \ , \ee and, equivalently, for $r(x)$
\be
\label{relationrx} \big(\frac{dr}{dx}\big )^2=
\frac{e^{2\rho}}{1+2 r \rho_r\ + \lambda \rho_r^2}\ . \ee

In the classical limit ($\lambda = 0$) Eqs. (\ref{eq:cg-7}) and
(\ref{relationrx}) become \be \label{classicallimiteq}
\rho_{rr}=-\frac{2(1+r\rho_r)\rho_r}{r}
 \  \end{equation}
 and
\be \label{relationrxclassical} \big(\frac{dr}{dx}\big )^2=
\frac{e^{2\rho}}{1+2 r \rho_r\ }\ .  \ee The general solution to
Eq. (\ref{classicallimiteq}) is \be \rho_{c} = \frac{1}{2}\ln
(A+\frac{B}{r}) \ee where $A$ and $B$ are two integration
constants. The Schwarzschild metric can be easily recovered by
setting $A=1$ (i.e., the metric is asymptotically Minkowskian,
$\rho\to 0$ as $r \to +\infty$) and
$B=-2GM/c^2$. Integration of (\ref{relationrxclassical}) leads to
the identification of $x$ with the tortoise coordinate $r^*$.


A natural thing would be to try to solve Eq. (\ref{eq:cg-7})
perturbatively in $\lambda$. This gives a good approximation to
the full solution when the quantum terms are small compared to the
classical ones. At $O(\lambda)$ this is true for large $r$ where
\be \lambda\rho_{r}\sim \lambda\frac{d\rho_{c}}{dr}= \frac{\lambda
GM/c^2 r^2}{1-2GM/c^2 r}\ll 2r \ , \ee but not in the near-horizon
region where the quantum terms instead dominate. In this region
backreaction effects are strong and cannot be treated
perturbatively.

\subsection{Numerical solution}
  The differential equation (\ref{eq:cg-7}) cannot be  solved  analytically.
 It must be  studied  numerically and for this
 it is  convenient to rescale the radial coordinate and
introduce the dimensionless parameter $z\equiv r/\sqrt{\lambda}$.
We then get
\begin{equation}\label{basicdeq}
\rho_{zz}=-\frac{(1+z\rho_z)\rho_z(2z+\rho_z)}{z^2-1}.
\end{equation}
This equation allows to analyze in a non-perturbative way the
exact function
 $\rho (z)$. However, for reasons that will be clear in a moment,
 the function $ \rho=\rho(z)$ is not single-valued. Therefore we have
to study, instead, the function $z=z(\rho )$ which verifies the
differential equation  \be \label{invfunc}
z_{\rho\rho}=\frac{(z+z_\rho)(1+2zz_\rho)}{z^2-1}\ . \ee

Imposing as boundary condition  that the solution behaves, for
very large $z$, as the classical one \be\label{condin} \rho
(z\rightarrow\infty)=\frac{1}{2}\ln(1-\frac{2a}{z}) \ee with
$a\equiv GM/c^2\sqrt{\lambda}$, we can generate numerically the
solution to (\ref{invfunc}). We find that the quantum corrected
solution is everywhere similar to the classical one, up to the
vicinity of the classical horizon. We can observe this behavior in
Fig.1. We have chosen a black hole of  small-size ($a=10^3$) since
in this case the differences between the classical and the
semiclassical solutions can be better appreciated.

\begin{figure}[htbp]
\begin{center}
\includegraphics[angle=0,width=3.5in,clip]{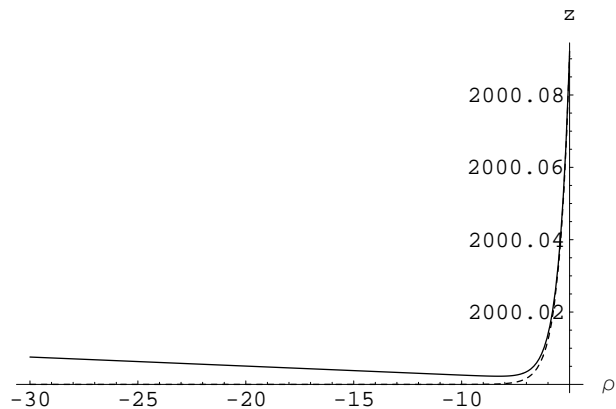}
\caption{Classical (dashed line) and numerical (solid line) plots
of the function $z(\rho)$ for $a=10^3$. }\label{Fig1}
\end{center}
\end{figure}

We observe that for regions far away from the classical horizon
($z >> 1$, i.e. $\rho\to 0$) the numerical solution and the
classical one are very similar. However, in the vicinity of the
classical horizon the quantum corrected solution  suffers a
bounce, absent in the classical solution, around $\rho \sim -8.3$
and then grows up slowly. This is the reason why we have had to
solve numerically $z=z(\rho)$ instead of $\rho=\rho(z)$. Note that
this point appears at a finite value of $\rho$, and, therefore,
that the time-time component of the metric $g_{tt}$  does not
vanish. The existence of this bouncing point is better represented
in Fig.2, where we plot the derivative $z_{\rho}$ in terms of
$\rho$. The existence of a zero for $z_{\rho}$ signals a bouncing
point for the radial function.

\begin{figure}[htbp]
\begin{center}
\includegraphics[angle=0,width=3.5in,clip]{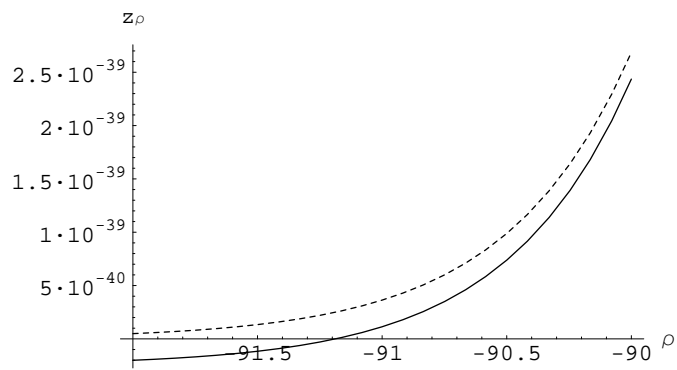}
\caption{Plots of $z_\rho(\rho)$, classical (dashed) and numerical
(solid) for $a=10^{39}$.} \label{Fig2}
\end{center}
\end{figure}

This qualitative behavior of the radial function is maintained
irrespective of the size of the black hole. For a Solar-mass black
hole ($a = 10^{39}$) the bounce  appears at $\rho(z_{B}) \sim
-91$. This result implies, as we will see later with more detail,
that the classical horizon is eliminated by the quantum
corrections. However, it is important to remark that the value of
the conformal factor of the metric at the bouncing surface is very
small. In fact the redshift for a signal emitted by a static
observer at the bounce and received at infinity \be
\frac{E_{\infty}}{E_{Bounce}} = e^{\rho(z_{B})} \  \ee transforms,
for instance, a Planckian energy $E_{Bounce}\sim 10^{19}GeV$ into
an energy of the order $E_{\infty}\sim 10^{-12} eV$ at infinity.
Moreover we find that, for $a=10^{39}$, $z_B = z_S + 8.48\times
10^{-4}$ , where $z_S\equiv 2a= 2 \times 10^{39}$. This shows that
the bouncing surface is indeed very close to the classical horizon
(the difference between $z_B$ and $z_S$ increases as one reduces
the mass).

\subsection{Behavior of the metric around the bounce}

Around the bounce $z_B$ the function $z(\rho)$ behaves \be z(\rho)
\approx z_{B}+\frac{1}{2}A(\rho - \rho_B)^2 + ... \ . \ee Plugging
this expansion into (\ref{invfunc}) we get immediately that \be
A=\frac{z_B}{z_B^2 -1} \ . \ee Therefore \be
\label{approximationrho} \rho(z) \approx \rho(z_B) \pm
\sqrt{\frac{2(z_B^2 -1)(z-z_B)}{z_B}} \ , \ee and \be
\label{approximaterhoz}\rho_z \approx \pm \sqrt{\frac{z_B^2
-1}{2z_B (z-z_B)}} \ . \ee

To estimate the form of the metric  at $r \sim r_B$ \be ds^2_{(4)}
\approx -e^{2\rho}c^2dt^2 + e^{2\rho}\left (\frac{dx}{dr}\right
)^2dr^2 + r^2d\Omega^2 \ \ee we need to find the relation between
$r$ and $x$. From Eq. (\ref{relationrx}) we have that \be
\label{exprelationrx} e^{2\rho}\left (\frac{dx}{dr}\right )^2=
1+2z\rho_z + \rho_z^2=1+2r\rho_{r}+\lambda\rho_{r}^2\ .   \ee In
the region $r\sim r_B$ the right hand side is dominated by the
pure quantum term $\rho_{z}^2=\lambda\rho_r^2$, with $\rho_z$
given in (\ref{approximaterhoz}), and so we have  \be \label{xr1}
r \approx
r_B+\frac{1}{2}\frac{r_B}{r_B^2-\lambda}e^{2\rho(z_B)}(x-x_B)^2 +
... \ \ . \ee Therefore the form of the metric is approximated by
\be \label{approxbounce}ds^2_{(4)} \approx -e^{2\rho}c^2dt^2 +
\frac{r_B^2 - \lambda}{2r_B^2}\frac{dr^2}{(1-\frac{r_B}{r})} +
r^2d\Omega^2
 \ . \ee
The quantum corrected geometry is not singular at the bounce, as
it can be checked that all curvature invariants are regular at
$r=r_B$.
Note that $r$ is not the good spatial coordinate to extend the
metric (\ref{approxbounce}) beyond $r_B$, one should rather use
$x$ (see (\ref{xr1})).

Finally we should briefly note that the surface $r=r_B $ is not an
event horizon since $g_{tt}(r_B)\neq 0$. However, there
$g_{rr}^{-1} (r_B)=0$ and this means that
$ \partial_+ r^2  < 0 $ for points $x$ inside  $r_B$ and only for
$r=r_B$ we have $
\partial_+ r^2 |_{r_B} = 0$.  This means that the surface $r=r_B$ still plays
the role of an apparent horizon for outgoing radiation.

\subsection{The geometry beyond the bounce}

We shall now investigate the geometry beyond the bouncing
surface. To this end we study the $2D$ curvature for $\rho <
\rho(z_B)$. Starting from the expression of the $2D$ curvature
\be \label{2dcurvaturex} R=8e^{-2\rho}\partial_+
\partial_- \rho=-2e^{-2\rho}\rho_{xx} \ , \ee where
$\rho_{xx}$ is given by (see Eq. (\ref{roxx}) where here
and in the next formulas $\dot\rho$ and $\ddot\rho$  are written
in terms of derivatives with respect to $z$)
 \be \label{rhoxx}
\rho_{xx}=(z\rho_z + z^2\rho_{zz})\phi_x^2 -z\rho_z \phi_{xx} \ ,
\ee and from (\ref{eq:cg-4}) we get  \be \label{2dcurvaturez}R=
-2z^2\frac{z\rho_{zz}+(2z+\rho_z)\rho_z^2}{z+\rho_z}e^{-2\rho}\phi_x^2
\ . \ee We can get a simplified expression for the curvature
taking into account our basic differential equation
(\ref{basicdeq}) for $\rho(z)$  \bea R &=&
-2z^2\frac{\rho_{zz}}{1+z\rho_z}e^{-2\rho}\phi_x^2 \nonumber \\
&=& 2z^2 \frac{\rho_z(2z+\rho_z)}{z^2-1}e^{-2\rho}\phi_x^2 \ .
\eea Moreover from  (\ref{relationrx})  we have \be
\label{expphix2} e^{-2\rho}\phi_x^2= \frac{1}{\lambda
z^2(1+2z\rho_z + \rho_z^2)} = \frac{z_{\rho}^2}{\lambda
z^2(1+2zz_{\rho} + z_{\rho}^2)} \ , \ee and therefore the final
expression for the curvature is  \be \label{2dcurvature}R=
\frac{2}{\lambda (z^2-1)}\frac{\rho_z(2z+ \rho_z)}{(1+2z\rho_z +
{\rho_z}^2)}=\frac{2}{\lambda (z^2-1)}\frac{(2zz_\rho+
1)}{(1+2zz_\rho + z_\rho^2)}  \ . \ee

The natural singularity at $z=1$ ($r= \sqrt{\lambda}$), which
mimics the classical singularity at $r=0$, does not belong to the
physical spacetime because of the existence of the bounce $r_B$
encountered before.
Singularities can only  be generated by zeroes of the denominator
in (\ref{2dcurvature}).  Before $r_B$ we have $z_{\rho}>0$, so
$1+2zz_\rho + z_\rho^2$  can never be zero. At $z=z_B$  we have
$z_{\rho}=0 $, so \be R= \frac{2}{\lambda (z_B^2-1)} \ . \ee After
the bounce  $z_{\rho}<0$ and, therefore, one can potentially
encounter a curvature singularity. The numerical analysis
indicates that such a  singularity can be found only when $ z \to
+\infty$. For this we need that $(1+2zz_\rho + z_\rho^2)\sim 0$
when $ z \to +\infty$. The zeroes of the above second order
polynomial $z_{\rho}= -z \pm \sqrt{z^2 -1}$ are an {\it exact}
solution to the differential equation (\ref{invfunc}). Since  the
numerical analysis shows that, for large $z$ in the interior
region, $z_{\rho} \sim -1/2z +... $ this means that asymptotically
we have $z_{\rho} \sim -z + \sqrt{z^2 -1} + f(z)$, where
$z^nf(z)\to 0$ as $z\to +\infty$ for every positive integer number
$n$.\footnote{We note that the
 solution $z_{\rho}=-z+\sqrt{z^2-1}$ corresponds to the one found
 in
 \cite{lowe} as an exact solution to Eqs. (\ref{eq:cg-i}), (\ref{eq:cg-ii})
 and (\ref{eq:cg-iii}) where the exponential term $e^{2(\phi+\rho)}/\lambda$
 is neglected.}
 The behavior of the curvature is
then \be R \sim -\frac{1}{ z^5 f(z)} \ , \ee which clearly shows
the existence of a singularity at $z=+\infty$. A detailed
computation, using (\ref{eq:cg-4}), shows that (for ${z \to
+\infty}$) \be \label{branchingpolyakov} z\sim\sqrt{-x} \ . \ee
Using Eq. (\ref{exprelationrx}) it is easy to realize that in the
limit $z\to +\infty$ we have $f(z)\sim  e^{-2z^2}$ and this
implies that the scalar curvature goes to $-\infty$ there as \be
R\sim -\frac{e^{2z^2}}{z^5} \ . \ee Moreover, such a singularity
is null (i.e. $e^{2\rho}\to 0$ as $z\to +\infty$) and is located
at a finite affine distance from any finite $z$. Finally we remark
that the singularity  arises due to the the branching point for
the radial function $r\equiv z \sqrt{\lambda}$, with respect to
the spatial coordinate $x$, displayed in
(\ref{branchingpolyakov}). This is the underlying reason for the
generation of the curvature singularity at $x=-\infty$.

\section{Quantum corrections in the $s$ wave approximation.}

The approximation used in the previous section consists,
essentially, in neglecting the effects of the potential barrier
for the wave equation (\ref{waveequations=0}). In this way we have
simplified considerably the technical problem. It is natural at
this point to ask  whether  the results  obtained are maintained
when the effects of the potential are included. The detailed
analysis presented before has allowed to introduce all the
techniques that we shall use to attack the full problem in the
s-wave approximation. The equations to solve are now more
involved. However, since the conceptual line to follow should be
clear we will focus only on the most important points.

In the Schwarzschild spacetime the expectation values of the
stress tensor components in the Boulware state (see
(\ref{uno})-(\ref{condboul}) and (\ref{condphi})) are \bea
\label{schbouldil} \langle B|T_{\pm\pm}| B\rangle &=&
\frac{\hbar}{24\pi} [ -\frac{2r_S}{r^3}+
\frac{15}{8}\frac{r_S^2}{r^4} ]
+\frac{\hbar}{16\pi r^2}(1-\frac{r_S}{r})^2\ln(1-\frac{r_S}{r})\ , \nonumber \\
\langle B|T_{+-}| B\rangle &=& \frac{\hbar}{12\pi}
[1-\frac{r_S}{r}]
\frac{r_S}{2r^3}\ , \nonumber \\
\langle\Psi| \frac{\delta S_m}{\delta \phi}|\Psi\rangle &=&
-\frac{7\hbar}{16\pi}\frac{r_S}{r^3} +
\frac{\hbar}{2\pi}\frac{r_S^2}{r^4}+\frac{\hbar}{8\pi
r^2}(1-\frac{2r_S}{r}) (1-\frac{r_S}{r})\ln(1-\frac{r_S}{r})\ . \
\ \ \ \  \ \ \ \eea All these quantities vanish asymptotically,
while on the horizon the leading divergence is the same as in the
Polyakov case (\ref{div}). Again, this means that to inspect the
near-horizon region $r\sim r_S$ we need to solve exactly the
backreaction equations.

\subsection{Decoupling the semiclassical equations}

We shall now proceed in parallel to Section $3$ to decouple the
system of differential equations (\ref{eq:cg-1})-(\ref{eq:cg-3})
to generate a single equation for $\rho=\rho(r)$. Using the
relations $\rho_x =\dot{\rho }\phi_x $ and $
\rho_{xx}=\ddot{\rho}(\phi_x)^2+\dot{\rho}\phi_{xx}$ we transform
them  into
\begin{eqnarray}
\phi_{xx}-\phi_x^2-2\dot{\rho}\phi_x^2&=&e^{2\phi}\left[\ddot{\rho}\phi_x^2+\dot{\rho}\phi_{xx}-\dot{\rho}^2\phi_x^2
+6\dot{\rho}\phi_x^2+6\rho\phi_x^2\right] \label{eq:cg-11}\\
\phi_{xx}-2\phi_x^2+\frac{e^{2(\phi+\rho )}}{\lambda}&=& e^{2\phi}\left[\ddot{\rho}\phi_x^2+\dot{\rho}\phi_{xx}
-3(\phi_{xx}-\phi_x^2)\right]  \label{eq:cg-21}\\
\phi_{xx}-\phi_x^2-\ddot{\rho}\phi_x^2-\dot{\rho}\phi_{xx}&=&e^{2\phi}[3\ddot{\rho}\phi_x^2+3\dot{\rho}\phi_{xx}+
6\dot{\rho}\phi_x^2+6\rho\phi_{xx} \nonumber \\
&+& \frac{3}{2}\frac{(\phi_{xx}-\phi_x^2)_x}{\phi_x}]. \ \ \ \ \ \
\label{eq:cg-31}
\end{eqnarray}

From (\ref{eq:cg-11}) and (\ref{eq:cg-21}) we get \bea
\phi_{xx}(1-\dot{\rho}e^{2\phi})&=&\left[1+2\dot{\rho}+e^{2\phi}(\ddot{\rho}-\dot{\rho}^2+6\dot{\rho}+6\rho)\right]\phi_x^2
\\ \phi_{xx}(1-(\dot{\rho}-3)e^{2\phi})&=& \left[ 2+
(\ddot{\rho}+3)e^{2\phi}\right ]\phi_x^2 - \frac{e^{2(\phi+\rho
)}}{\lambda} \ , \eea which can be rewritten as \bea
\label{identityphix2} \phi_x^2&=&\frac{(1-\dot{\rho}e^{2\phi})
e^{2(\phi+\rho)}}{\lambda D} \ , \\
\label{identityphixx}\phi_{xx}&=&\frac{\left [
1+2\dot{\rho}+e^{2\phi}(\ddot{\rho}-\dot{\rho}^2+6\dot{\rho}+6\rho)\right
]e^{2(\phi+\rho)}}{\lambda D} \ , \eea where \bea D = &-&\left [
1-(\dot{\rho}-3)e^{2\phi}\right ]\left
[1+2\dot{\rho}+e^{2\phi}(\ddot{\rho}-\dot{\rho}^2+6\dot{\rho}+6\rho)\right]
 \nonumber \\
&+&\left [ 1-\dot{\rho}e^{2\phi} \right ] \left [2+
(\ddot{\rho}+3)e^{2\phi}\right ] \ . \eea Taking into account that
\be \frac{\phi_{xxx}}{\phi_x}= \dot{\phi}_{xx} \ , \ee and
plugging the expressions (\ref{identityphix2}) and
(\ref{identityphixx}) into (\ref{eq:cg-31}) we obtain,  after a
straightforward but very long calculation,  a third-order
differential equation relating $\rho$ and $r$

\bea \label{exactdifferentialequation}
&&\bigg[r^2+3\lambda-6\lambda\rho+r\rho_r(2r^2+6\lambda+r\lambda\rho_r)\bigg]\times
\bigg[72\lambda^2\rho^2(r^2+3\lambda+r\lambda\rho_r)+ \nonumber \\
&&+6r\lambda\rho\bigg(12r
\lambda-\rho_r(6r^4+40r^2\lambda+63\lambda^2+2r\lambda
\rho_r(4r^2+14\lambda+r\lambda\rho_r))+\nonumber\\
&&+r\lambda(2r^2+15\lambda+2r\lambda\rho_r)\rho_{rr}\bigg)+r^2\bigg(r\lambda(6r^4+56r^2\lambda+93\lambda^2)\rho_r^3+\nonumber\\
&&+2r^2\lambda^2(r^2+6\lambda)\rho_r^4+\rho_r^2(4r^6+54r^4\lambda+162r^2\lambda^2+162\lambda^3-9r^2\lambda^3\rho_{rr})+\nonumber\\
&&+r^2(2(r^4+5r^2\lambda+12\lambda^2)\rho_{rr}+9\lambda^3\rho_{rr}^2-3r\lambda^2\rho_{rrr})
+r\rho_r(4r^4-6r^2\lambda- \nonumber
\\ &&72\lambda^2+\lambda(2r^4-14r^2\lambda-69\lambda^2)\rho_{rr}-3r\lambda^3\rho_{rrr})\bigg)\bigg]=0
 \eea
 The second factor in the above equation is the relevant one since
 it leads directly, when $\lambda =0$, to the classical equation
 (\ref{classicallimiteq}). The quantum corrections to the Schwarzschild metric should be
 then computed by exactly solving the above differential equation. Introducing the dimensionless coordinate
$z=e^{-\phi}=r/\sqrt{\lambda}$ we  get an ordinary differential
equation for $\rho=\rho(z)$. However, as we have already explained
in the analysis of the Polyakov theory, since the function
$\rho=\rho(z)$ could not be one-to-one it is more appropriate to
work directly with the differential equation for the function
$z=z(\rho)$. It reads as follows \bea \label{invfunctiondilaton}
&&18\rho(-21+4\rho)z z_\rho^4+216\rho^2z_\rho^5+2z^8
z_\rho^2(2z_\rho-z_{\rho\rho})+2z^7z_\rho(3z_\rho+2z_\rho^3
\nonumber \\&&-z_{\rho\rho})
+3z^3z_\rho\bigg((31-4\rho)z_\rho-8(3+10\rho)z_\rho^3+(23-4\rho)z_{\rho\rho}\bigg) \nonumber \\
&&+6z^2z_\rho^2\bigg((27-28\rho)z_\rho+12\rho(1+\rho)z_\rho^3-15\rho
z_{\rho\rho}\bigg)+2z^6z_\rho(1+27z_\rho^2 \nonumber \\ &&-5z_\rho
z_{\rho\rho})
-3z^4\bigg(-3z_{\rho\rho}+2z_\rho(-2+(-27+8\rho)z_\rho^2+2(2+\rho)z_\rho
z_{\rho\rho})\nonumber \\&&-z_{\rho\rho\rho}\bigg)
+z^5\bigg(56z_\rho^2-6(1+6\rho)z_\rho^4-9z_{\rho\rho}^2+z_\rho(14z_{\rho\rho}+3z_{\rho\rho\rho})\bigg)=0
\ . \ \ \ \eea

\subsection{Numerical solution}

\begin{figure}[htbp]
\begin{center}
\includegraphics[angle=0,width=3.5in,clip]{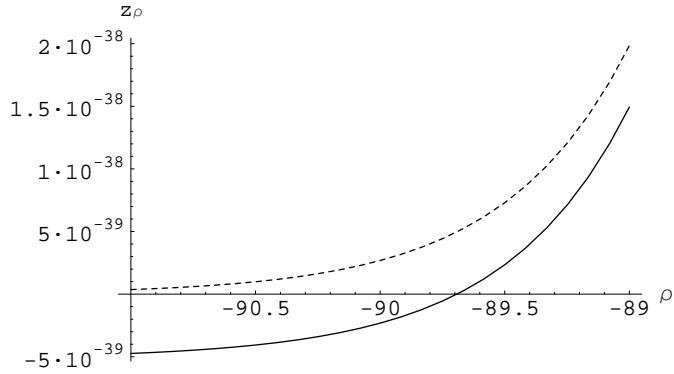}
\caption{Plots of the function $z_\rho(\rho)$, classical (dashed
line) and numerical (solid line) for
$a=10^{39}$.}\label{dilatonder}
\end{center}
\end{figure}

We solve numerically the above equation by imposing that, for very
large $z$ and $\rho \to 0$, the solution approaches the classical
one. We find that the solution is almost identical to the
classical one up to the vicinity of a surface, which we also
denote as $z_B$, located very close to the classical horizon. The
result (for a Solar-mass black hole $ a\equiv GM/c^2\sqrt{\lambda}
=10^{39}$) is depicted  in Fig. \ref{dilatonder}
which shows  the existence of a bounce   for the radial function
at $\rho_B \approx -89.69$, where \be z(\rho) \approx
z_{B}+\frac{1}{2}A(\rho - \rho_B)^2 + ... \ , \ee and $A$ is a
positive  coefficient to be computed numerically. Moreover, we
find that $z_B \approx z_S + 1.76 \times 10^{-37}$. Therefore, its
location is much closer to the classical horizon than in the
Polyakov theory approximation of section 3.

\subsection{Branching point for the radial function}

\begin{figure}[htbp]
\begin{center}
\includegraphics[angle=0,width=3.5in,clip]{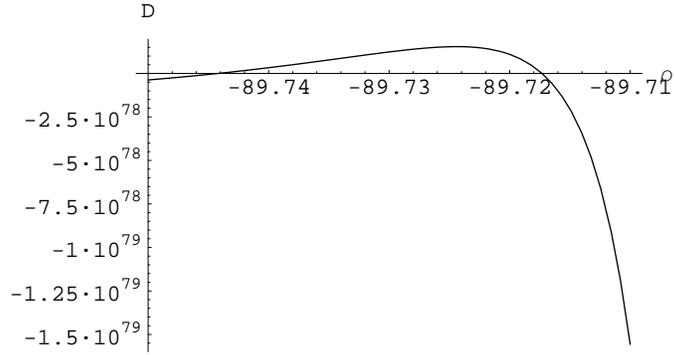}
\caption{For $a=10^{39}$ the bounce  of the radial function
($D=+\infty$) is located at $\rho\sim -89.69$.  At $\rho\sim
-89.72$  the function $D$ vanishes.}\label{bouncingtortoise}
\end{center}
\end{figure}

The main difference with respect to the Polyakov theory appears in
the relation between $r$ and  $x$. In terms of the function
$z(\rho)$ we have \bea
e^{2\rho}(\frac{dx}{dr})^2&=&\frac{zz_{\rho}D}{(1+zz_{\rho})} \\
e^{2\rho}(\frac{dx}{d\rho})^2&=&\frac{\lambda
zz_{\rho}^3D}{(1+zz_{\rho})} \eea
 where \bea \label{D} D=\frac{1}{z^4z_{\rho}^4}
\{&-&[z^2z_{\rho}+z+3z_{\rho}][z^2z_{\rho}^3-2z^3z_{\rho}^2-z^2z_{\rho\rho}
-5zz_{\rho}^2-z^2z_{\rho}+6\rho z_{\rho}^3]\nonumber \\
&+&[z^2z_{\rho}+z][2z^2z_{\rho}^3+3z_{\rho}^3 -
z^2z_{\rho\rho}+zz_{\rho}^2]\} \ . \eea In Fig.
(\ref{bouncingtortoise}) we show the typical behavior of the
function $D$ in terms of $\rho$.

Note that in the vicinity of the bounce $z_\rho\rightarrow 0$, and
from (\ref{D}) we have $ D\sim z_{\rho}^{-3}$. Therefore \be
e^{2\rho}(\frac{dx}{dr})^2\sim\frac{1}{z_\rho^2} \ , \ee and hence
\be \frac{dr}{dx}\sim z_{\rho} \ . \ee In other words, the
expansion of $r$ in terms of $x$ has to be of the form $r\approx
r_B+ \alpha (x-x_B)^2 + ... $, where $\alpha$ is a numerical
constant, in agreement with the analytic behavior encountered in
(\ref{xr1}).

A similar argument allows to determine the behavior of $r$ in
terms of $x$ around the zero of the function $D$. This happens at
$\rho=\rho_M  \stackrel{<}{\sim} \rho_{B}$, just after the bounce.
Around the zero of $D$ we have $D(r)\sim r_M-r$. Therefore \be
\frac{dx}{dr}\sim-\sqrt{r_M-r} \ , \ee and hence
\be\label{rbranching} r\approx r_M-\beta(x-x_M)^{2/3}\ee where
$\beta$ is a numerical positive constant. The radial function has
a branching point at $x=x_M$, which turns out to be the minimum
possible value for the ``tortoise'' coordinate $x$.

The form of the metric in this region is
 \be \label{metricaM}ds^2_{(4)} \approx
e^{2\rho(r)}(-c^2dt^2+ dx^2)+(r_M+B(x-x_M)^{2/3})^2d\Omega^2 \ee
where, according to our previous analysis, the function $\rho(r)$
is finite and regular at $r_M$. The above metric has a singularity
at $r=r_M$, which cannot be avoided by a change of coordinates. It
is indeed a curvature singularity as we now show. The $4D$ scalar
curvature can be expressed, in terms of $\rho$ and $\phi$, as
follows
\begin{equation} \label{4dcurvature}
R^{(4)}= -2e^{-2\rho
}\left[\rho_{xx}-2\phi_{xx}+3\phi_x^2\right]+2e^{2\phi} \ .
\end{equation}
The first term is just the two-dimensional scalar curvature
$R=-2e^{-2\rho }\rho_{xx}$, which according to  (\ref{rhoxx}),
(\ref{identityphix2}) and (\ref{identityphixx}) is  \be
R=-\frac{2}{\lambda
z^2D}\frac{-z^3z_{\rho\rho}+2z^3z_{\rho}+z^2+6zz_{\rho}-6\rho
z_{\rho}}{zz_{\rho}^3} \ . \ee At $x=x_M$, where $D(x_M)=0$, $R$
and also the second and third terms in (\ref{4dcurvature}) are
divergent. Finally we note, from (\ref{metricaM}), that this
singularity is timelike, has finite radius $r_M$ and is located at
a finite affine distance away.

\section{Conclusions}

The existence of a bounce for the radial function $r$, prior to
the emergence of a spacetime singularity, is perhaps  the most
significative result of our analysis.
It already appears in the simplified Polyakov theory and it is
still there in the most accurate $s$-wave approximation. It is
natural to expect it to persist in a full treatment of the
problem.

Our results
 are perhaps not surprising from the semiclassical point of view, where
due to the strong divergence of the Boulware stress tensor at the
Schwarzschild horizon important deviations from the classical
behavior are indeed expected to arise. However they constitute an
important prediction for the search of static 5D braneworld
configurations with asymptotically flat boundary conditions on the
brane: such solutions are not black holes, but rather naked
singularities. In the 4D semiclassical context this reinforces the
idea that the Boulware state describes the vacuum polarization
around a static star, not a black hole. Indeed the natural thing
for a black hole is to be time dependent and to evaporate via the
Hawking effect.

Finally we point out  that our analysis does not exclude the
existence of static braneworld black holes, but the price to pay
is to give up asymptotic flatness on the brane. This means, in the
dual $4D$ theory, to replace Boulware with the Hartle-Hawking
state. \\

\noindent {\bf Acknowledgements}\noindent

We thank R. Balbinot, S. Fagnocchi and G.P. Procopio for useful
discussions and J.M. Martin-Garcia, A. Nagar, M. Nebot and C.
Talavera for assistance in the numerical analysis.


\begin{thebibliography}{99}
\bibitem{hawk1} S. W. Hawking,  {\it Comm. Math. Phys.} {\bf 43}
199 (1975)
\bibitem{parker}L. Parker  {\it Phys. Rev.} D {\bf 12} 1519 (1975)
\bibitem{wald} R. M. Wald {\it Comm. Math. Phys.} {\bf 45} 9
(1975)
\bibitem{hawk2} S.W. Hawking, {\it Phys. Rev.} {\bf D14}, 2460 (1976)

\bibitem{birreldavies} N. D. Birrel and P. C. W. Davies  {\it
Quantum fields in curved spaces}, Cambridge University Press,
Cambridge (1982)

\bibitem{frolovnovikov} V. P. Frolov  and I. D. Novikov  {\it Black
hole physics}, Kluwer Academic Publisher, Dordrecht (1996)

\bibitem{icp05} A. Fabbri  and J. Navarro-Salas  {\it Modeling black hole evaporation},
Imperial College Press-World Scientific, London (2005)



\bibitem{unit} See for instance G.'t Hooft, {\it Nucl. Phys.} {\bf B335}, 138 (1990);
T. Banks, A. Dabholkar, M.R. Douglas and M. O'Loughlin, {\it Phys.
Rev.} {\bf D45}, 3607 (1992); K. Schoutens, H.L. Verlinde and E.P.
Verlinde, {\it Phys. Rev.} {\bf D48}, 2670 (1993); L. Susskind, L.
Thorlacius and J. Uglum, {\it Phys. Rev.} {\bf D48}, 3743 (1993);
C.R. Stephens, G. 't Hooft and B.F. Whiting, {\it Class. Quant.
Grav.} {\bf 11}, 621 (1994); D.A. Lowe, J. Polchinski, L.
Susskind, L. Thorlacius and J. Uglum, {\it Phys. Rev.} {\bf D52},
6997 (1995); R.C. Myers, {\it Gen. Rel. Grav.} {\bf 29}, 1217
(1997); D. Amati, {\it J. Phys.} {\bf G29}, 31 (2003); G.T.
Horowitz and J. Maldacena, {\it JHEP} {\bf 0402}, 008 (2004); A.
Ashtekar and M. Bojowald, {\it Class. Quant. Grav.} {\bf 22}, 3349
(2005)
\bibitem{hawk3} S.W. Hawking, {\it Information loss in black holes}, hep-th/0507171
\bibitem{boulware} D. Boulware, {\it Phys. Rev.} {\bf D11}, 1404 (1975)
\bibitem{chrisful} S.M. Christensen and S.A. Fulling, {\it Phys. Rev.}{\bf D15},
2088 (1977)
\bibitem{harhaw} J.B. Hartle and S.W. Hawking, {\it Phys. Rev.}{\bf D13}, 2188 (1976);
W. Israel, {\it Phys. Lett.}{\bf A57}, 107 (1976)
\bibitem{unruh} W.G. Unruh, {\it Phys. Rev.}{\bf D14}, 870 (1976)

\bibitem{ransun} L. Randall and R. Sundrum, {\it Phys. Rev. Lett.} {\bf 83}, 4690 (1999)

\bibitem{malda} O. Aharony, S. Gubser, J. Maldacena, H. Ooguri and Y. Oz {\it Phys. Rept.}
{\bf 323}, 183 (2000)



\bibitem{molti} A. Chamblin, S.W. Hawking and H.S. Reall, {\it Phys. Rev.} {\bf D61},
065007 (2000); N. Dadhich, R. Maartens, P. Papadopoulos and V. Rezania,
{\it Phys. Lett.} {\bf B487}, 1 (2000); 
M. Bruni, C. Germani and R. Maartens, {\it Phys. Rev. Lett.} {\bf 87},
231302 (2001);
P. Kanti and K. Tamvakis,
{\it Phys. Rev.} {\bf D65}, 084010 (2002); R. Casadio, A. Fabbri and
L. Mazzacurati, {\it Phys. Rev.} {\bf D65}, 084040 (2002); G. Kofinas,
E. Papantonopoulos and V. Zamarias, {\it Phys. Rev.} {\bf D66}, 104028 (2002);
H. Kudoh, T. Tanaka and T. Nakamura, {\it Phys. Rev.} {\bf D68}, 024035 (2003);
C. Charmousis and R. Gregory, {\it Class. Quant. Grav.} {\bf 21}, 527 (2004)

\bibitem{taefk} R. Emparan, A. Fabbri and N. Kaloper, {JHEP} {\bf 08}, 043 (2002);
T. Tanaka, {\it Prog. Theor. Phys. Suppl.} {\bf 148}, 307 (2003)

\bibitem{abf} P.R. Anderson, R. Balbinot and A. Fabbri, {\it Phys. Rev. Lett.} {\bf 94},
061301 (2005); J. Garriga and T. Tanaka, {\it Phys. Rev. Lett.}
{\bf 84}, 2778 (2000); M.J. Duff and J.T. Liu, {\it Phys. Rev.
Lett.} {\bf 85}, 2052 (2000)
\bibitem{ahs} P.R. Anderson, W.A. Hiscock and D.A. Samuel, {\it Phys. Rev.} {\bf D51},
4337 (1995)

\bibitem{lowe} D.A. Lowe, {\it Phys. Rev.} {\bf D47}, 2446 (1993)



\end{thebibliography}
\end{document}